\RequirePackage{fix-cm}

\documentclass[twocolumn]{svjour3}

\usepackage{graphicx}
\usepackage{url}
\usepackage{color}
\usepackage{paralist}
\usepackage{fancyvrb}
\usepackage{epsfig}
\usepackage{subfig}
\usepackage[numbers]{natbib}
\usepackage{tabularx}
\usepackage[table,xcdraw]{xcolor}
\usepackage{epstopdf}
\usepackage[subtle, tracking=normal,leading=tight]{savetrees}
\usepackage{multirow}
\usepackage{soul}
\usepackage{booktabs}
\usepackage{multirow}

\newcommand{\q}[1]{\lq\lq{}{}#1\rq\rq{}{}}
\newcommand{\tool}[0]{{\em ArchiveSpark2Triples}}

\begin{document}\sloppy

\title{Building and Querying Semantic Layers for Web Archives (Extended Version)\footnote{\thanks{* This
is an extended version of the paper: {\em P. Fafalios, H. Holzmann,
V. Kasturia, \& W. Nejdl, \lq\lq{}{}Building and Querying Semantic
Layers for Web Archives\rq\rq{}{}, 2017 ACM/IEEE-CS Joint Conference
on Digital Libraries, June 2017.}}}}

\titlerunning{Building and Querying Semantic Layers for Web Archives (Extended Version)}

\author{Pavlos Fafalios \and
        Helge Holzmann  \and
        Vaibhav Kasturia \and
        Wolfgang Nejdl
       }

\institute{Pavlos Fafalios  \and
           Helge Holzmann   \and
           Vaibhav Kasturia \and
           Wolfgang Nejdl
              \at
              L3S Research Center, Leibniz University of Hannover,
              Appelstr. 9a, 30167 Hannover, Germany\\
              \email{\{fafalios, holzmann, kasturia, nejdl\}@L3S.de}
          }

\date{Received: date / Accepted: date}

\maketitle

\begin{abstract}
Web archiving is the process of collecting portions of the Web to
ensure that the information is preserved for future exploitation.
However, despite the increasing number of web archives worldwide,
the absence of efficient and meaningful exploration methods still
remains a major hurdle in the way of turning them into a usable and
useful information source. In this paper, we focus on this problem
and propose an RDF/S model and a distributed framework for building
semantic profiles (\q{layers}) that describe semantic information
about the contents of web archives. A semantic layer allows
describing metadata information about the archived documents,
annotating them with useful semantic information (like entities,
concepts and events), and publishing all this data on the Web as
Linked Data. Such structured repositories offer advanced query and
integration capabilities, and make web archives directly exploitable
by other systems and tools. To demonstrate their query capabilities,
we build and query semantic layers for three different types of web
archives. An experimental evaluation showed that a semantic layer
can answer information needs that existing keyword-based systems are
not able to sufficiently satisfy.


\end{abstract}

\section{Introduction}
Significant parts of our cultural heritage are produced and consumed on the Web.
However, the ephemeral nature of the Web makes most of its information
unavailable and lost after a short period of time.
Aiming to avoid losing important historical information,
a web archive captures portions of the Web to ensure the information
is preserved for future researchers, historians,
and interested parties in general.

Despite the increasing number of web archives worldwide,
the absence of efficient and meaningful exploration methods
still remains a major hurdle in the
way of turning web archives into a usable and useful source of information.
The main functionalities offered by existing systems are to find
older versions of a specific web page,
to search on specific collections,
or to search using keywords
and filter the retrieved results by
selecting some basic metadata values.
However,
for a bit more complex information needs,
which is usually the case when exploring web archives,
keyword-based search
leads to ineffective interactions and poor results \cite{weikum2011longitudinal}.
This is true especially for {\em exploratory search} needs
where searchers
are often unfamiliar with the domain of their goals,
unsure about the ways to achieve their goals,
or need to learn about the topic in order to understand how to
achieve their goals \cite{marchionini2006exploratory}.
Thus, for exploring web archives,
there is the need to go beyond keyword-based search and
support more advanced information seeking
strategies \cite{weikum2011longitudinal,whitelaw2015generous,holzmannaccessing2017}.

To cope with this problem,
we propose building semantic profiles (\q{layers}) that describe semantic information about
the contents of archived documents.
Specifically, we base upon Semantic Web technologies and
propose an RDF/S \cite{brickley2014rdf} data model that allows:
a) describing useful metadata information about each archived document,
b) annotating each document with entities, concepts and events
extracted from its textual contents,
c) enriching the extracted entities\footnote{For simplicity,
when we say entity we refer to entity
(e.g., {\em Barack Obama}, {\em New York}, or {\em Microsoft}), concept (e.g., {\em Democracy} or {\em Abortion})
or event (e.g., {\em 2010 Haiti earthquake} or {\em 2016 US Election}).}
with more semantic information (like properties and related entities coming from
other knowledge bases), and
d) publishing all this data on the Web in the standard RDF format,
thereby making all this information directly accessible and exploitable by other systems and tools.
Then, we can use this model for creating and maintaining a semantic repository
of structured data about a web archive.
Note that the actual contents of the web archive are not
stored in the repository.
The proposed approach only stores metadata information
that allows identifying interesting documents and information based on
several aspects (time, entity, type or property of entities, etc.).
Therefore, such a repository acts as a {\em semantic layer} over the archived documents.

By exploiting the expressive power of SPARQL \cite{prud2008sparql}
and its federated features
\cite{prud2013sparql,fafalios2016querying}, we can run advanced
queries over a semantic layer. For example, in case we have
constructed a semantic layer for a {\em news archive}, we can run
queries like:

\begin{itemize}
\item find articles of 1995 discussing about New York lawyers
\item find medicine-related articles published during 1995
\item find out the most discussed politician during 1995
\item find out politicians discussed in articles of 1990 together with {\em Nelson Mandela}
\item find out how the popularity of {\em Barack Obama}~ evolved over time during 2007
\item find articles similar to another article
\end{itemize}

Note that for all these queries we can directly (at query-execution time)
integrate information coming from online knowledge bases like DBpedia \cite{lehmann2015dbpedia}.
For instance, regarding the first query, for each lawyer
we can directly access DBpedia and
retrieve his/her birth date,
a photo and a description in a specific language.
Thus, semantic layers enable connecting
web archives with existing knowledge bases.

In a nutshell, in this paper we make the following contributions:
\begin{itemize}
\item   We introduce a simple but flexible RDF/S data model,
        called {\em Open Web Archive},
        which allows describing and publishing metadata and semantic information about
        the contents of a web archive.
\item   We detail the process
        of constructing semantic layers and
        we present an open source and distributed framework,
        called \tool, that
        facilitates their efficient construction.
\item   We present (and make publicly available)
        three semantic layers for three different types of web archives:
        one for a {\em  versioned web archive},
        one for a {\em non-versioned news archive},
        and one for a {\em social media archive}.
\item   We showcase the query capabilities offered by semantic layers
        through many interesting exploitation scenarios and query examples.
\item   We detail the results of a comparative evaluation using a set
        of 20 information needs of exploratory nature
        (providing also their relevance judgements).
        The results showed that a semantic layer can satisfy information
        needs that existing keyword-based systems
        are not able to sufficiently satisfy. They also enabled
        us to identify problems that can affect the effectiveness of
        query answering.
\end{itemize}

The rest of this paper is organized as follows:
Section \ref{sec:motivAndRW} motivates our work and presents
related literature.
Section \ref{sec:constrSemLayers} introduces the
{\em Open Web Archive} data model and describes
the process and a framework for constructing semantic layers.
Section \ref{sec:casestudies} presents three semantic layers
for three different types of web archives, as well as
their query capabilities.
Section \ref{sec:eval} presents evaluation results.
Finally, Section \ref{sec:concl} concludes the paper and
discusses directions
for future research.

\section{Motivation and Related Work}
\label{sec:motivAndRW}

In this section, we first
motivate our work by discussing information needs that our approach intends to
satisfy for enabling more sophisticated search and exploration of web archives.
Then we review related works by also discussing the difference of our approach.

\subsection{Motivation}
\label{motivation}

Working with large web archives in the context of the
ALEXANDRIA project\footnote{The ALEXANDRIA project (ERC Advance Grant, Nr. 339233, \url{http://alexandria-project.eu/})
aims to develop models, tools and techniques necessary to explore and analyze web archives in a meaningful way.},
we have identified the following information needs
that an advanced exploration system for web archives should satisfy:

\begin{itemize}
\item[Q1]   {\em Information Exploration.}
            How to explore documents about entities from the past in a
            more advanced and \q{exploratory} way,
            e.g., even if we do not know the entity names
            related to our information need?
            For example, how can we find articles of a specific time period
            discussing about a specific category of entities (e.g., {\em philanthropists})
            or about entities sharing some characteristics (e.g., {\em born in Germany before 1960})?
\item[Q2]   {\em Information Integration.}
            How to explore web archives by also integrating
            information from existing kno\-wledge bases?
            For example, how can we find articles discussing about
            some entities and for each entity to also retrieve
            and show some characteristics (e.g., an image or a description in a specific language)?
            Cross-domain knowledge bases like DBpedia contain such properties for almost every popular entity.
            Moreover, how to directly integrate information coming from
            multiple web archives?
            For example, how can we combine information
            from a news archive and a social media archive?
\item[Q3]   {\em Information Inference.}
            How to infer knowledge by exploiting the contents of a web archive? For example,
            can we identify important time periods related to one or more entities?
            Vice-versa, can we find out the most popular entities of a specific type
            in a specific time period (e.g., most discussed {\em politicians} in articles of {\em 2000})?
            Or how can we understand the topic of a web page (e.g., find news articles related to {\em medicine})?
\item[Q4]   {\em Robustness (in information change).} How to explore a web archive by automatically taking into account the change
            of entities over time? For example, the company {\em Accenture} was formerly known
            as {\em Andersen Consulting}, or
            the city {\em Saint Petersburg} was previously named {\em Leningrad}.
            Such temporal reference variants are common in the case of high impact events,
            new technologies, role changes, etc. How can we find
            documents from the past about such entities without having to worry about their correct reference?
\item[Q5]   {\em Multilinguality.} How to explore documents about
            entities from the past
            independently of the document language (and thus of the language
            of the entity name)? For instance,
            {\em abortion} is {\em Avortement} in French and
            {\em Sch\-wan\-ger\-schaftsabbruch} in German.
            How can we find documents about entities without having to worry
            about the document and entity language?
\item[Q6]   {\em Interoperability.}
            How to facilitate exploitation of web archives by other systems?
            How to expose information about web archives
            in a standard and machine understandable format,
            that will always be available on the Web,
            and that will allow for easy information integration?
            How to avoid downloading and parsing the entire web archive for
            identifying an interesting part of it related to a time period,
            some metadata values, and/or some entities. For example, how can we gather a corpus
            of articles of {\em 2004} discussing about {\em Greek politicians}?
\end{itemize}

\subsection{Related Work}
\label{rw}

Below we discuss related works on
{\em profiling}, {\em exploring}, and {\em analyzing} web archives,
and we discuss the differences and limitations of our approach.

\subsubsection{Profiling Web Archives}
A semantic layer can be considered a way to {\em profile}
the contents of a web archive.
AlSum et al. \cite{alsum2014profiling} exploit the age of the archived copies
and their supported domains,
to avoid sending queries to archives that likely do not hold the archived page.
Alam et al. \cite{alam2015web} examine the size and precision
trade-offs in different policies for producing profiles of web archives
(ranging between using full URIs and only top-level domains).
Bornand et al. \cite{bornand2016routing}
explore the use of binary, archive-specific classifiers
to determine whether or not to query an archive for a given URI.
Finally, Alam et al. \cite{alam2016web}
introduce a random searcher model to randomly explore the holdings of an archive
by exploiting the co-occurrence of terms.

\subsubsection*{Difference of our approach}

The aim of all these works is to improve the effectiveness of
query routing strategies in a distributed archive search environment.
However, such profiling approaches do not allow expressing semantic information about the
{\em contents} of the archived documents and thus cannot be exploited for satisfying more sophisticated
information needs like those discussed in Section \ref{motivation}.

\subsubsection{Exploring Web Archives}

\subsubsection*{Online services}
The Wayback Machine is a digital archive
of the Web
created by the Internet Archive\footnote{\url{https://archive.org}}.
It currently contains more than
450 billion web pages, making it the biggest web archive in the world.
With the Wayback Machine, the user can retrieve and access older versions of a web page.
The results are displayed in a calendar view
showing also the number of times the URL was crawled.
Wayback Machine also offers faceted exploration of archived collections,
thus allowing the user to filter the displayed results by media type, subject,
collection, creator, and language.
Recently, it also started offering keyword-based searching.

The Portuguese Web Archive (PWA)\footnote{\url{http://archive.pt}}
is a research infrastructure that enables search and access to files archived from the Web since 1996.
PWA provides comprehensive crawls of the Portuguese Web
and supports both keyword and URL based searching.

Memento's {\em Time Travel} service\footnote{\url{http://mementoweb.org}}
makes it easier for users
to browse the archived version of a web page
by redirecting them to the archive hosting the page.
The user provides the URL of the web page and a date of interest
and Time Travel checks various web archives for finding
an older version of the web page closest to the time indicated by the user.

Archive-It\footnote{\url{https://archive-it.org}} is a
web archiving service from the Internet Archive that helps
harvesting, building and preserving collections of digital content.
It currently supports keyword-based searching
while the user can also filter the displayed results based on several
metadata values like creator, subject, and language.
Padia et al.\cite{padia2012visualizing} present an alternative interface for
exploring an Archive-It collection
consisting of multiple visualizations (image plot with histogram, wordle,
bubble chart and timeline).

\subsubsection*{Research works}

Regarding research works,
Tempas \cite{holzmann2016tempas}
is a keyword-based search system that exploits a social bookmarking service for
temporally searching a web archive by indexing tags and time.
It allows temporal selections
for search terms, ranks documents based on their
popularity and also provides query recommendations.
The new version of Tempas \cite{holzmann2017exploring}
makes use of temporal link graphs and the corresponding anchor texts.
The authors show how temporal anchor texts can be effective in answering
queries beyond purely navigational intents,
like finding the most central web pages of an entity in a given time period.

Kanhabua et al. \cite{kanhabua2016search}
propose a search system that uses Bing for searching the current Web
and retrieving a ranked list of results.
The results are then linked to the WayBack Machine
thereby allowing keyword search on the Internet Archive without processing
and indexing its raw contents.

Vo et al. \cite{vo2016can}
study the usefulness of non-content evidences
for searching web archives,
where the evidences are mined only from metadata of the web pages,
their links and the URLs.

ArchiveWeb \cite{fernando2016archiveweb}
is a search system that supports collaborative search
of archived collections.
It allows searching across multiple collections
in conjunction with the live web,
grouping of resources,
and enrichment using comments and tags.

Jackson et al.\cite{jackson2016desiderata} present two prototype
search interfaces for web archives.
The first provides facets to filter the displayed results by several metadata values
(like content type and year of crawl), while
the other is a trend visualization inspired by Google's Ngram Viewer.

Singh et al.\cite{singh2016history}
introduce the notion of {\em Historical Query Intents}
and model it as a search result diversification task
which intends to present the most relevant results (for free text queries) from a
topic-temporal space.
For retrieving and ranking historical documents (e.g., news articles),
the authors propose a novel retrieval algorithm, called HistDiv,
which jointly considers the aspect and time dimensions.

Expedition \cite{singh2016expedition} is a time-aware search system for scholars.
It allows users to search articles in a news collection
by entering free-text queries and choosing from four retrieval models:
Temporal Relevance, Temporal Diversity, Topical Diversity, and Historical Diversity.
The results are presented in a
newspaper-style interface, while entity filters allows users refine the results.

The work by Matthews et al. \cite{matthews2010searching} proposes {\em Time Explorer},
an application designed to help users see how topics and entities associated
with a free-text query change over time.
By searching on time expressions
extracted automatically from text, Time Explorer
allows users to explore how topics evolved in the
past and how they will continue to evolve in the future.

\subsubsection*{Difference of our approach}

Although most of the existing approaches offer user-friendly interfaces,
they cannot satisfy more complex (but common) information needs
like those described in Section \ref{motivation}.
By basing upon semantic technologies, a semantic layer
allows to {\em semantically} describe the contents of a web archive
and to directly \q{connect} them with existing information available
on online knowledge bases like DBpedia.
In that way, we are able not only to explore archived documents in a more advanced way,
but also integrate information, infer new knowledge and quickly
identify interesting parts of a web archive for further analysis.

A similar approach to our work has been recently proposed by Page et al. \cite{page2017realising}.
In this work, the authors build a {\em layered} digital library
based on content from the Live Music Internet Archive.
Starting from the recorded audio and basic information
in the archive, this approach first deploys a {\em metadata layer} which
allows an initial consolidation of performer, song,
and venue information. A {\em processing layer} extracts audio features
from the original recordings, workflow provenance, and summary
feature metadata, while a {\em further analysis} layer provides tools for the
user to combine audio and feature data, discovered and reconciled
using interlinked catalogue and feature metadata from the other layers.
Similar to our approach, the resulting layered digital library
allows exploratory search across and within its layers.
However, it is focused on music digital libraries and
requires the availability of a large amount of metadata
which is not usually the case in web archives.
On the contrary, our approach focuses on {\em entity-centric} analysis and exploration of
an archived collection of documents.

The main drawback of our approach is its user-friendliness since,
currently, for querying a semantic layer one has to write structured (SPARQL) queries.
However, user-friendly interfaces can be developed on top of semantic layers
that will allow end-users to easily explore them.
Moreover, we can directly exploit
systems like Sparklis \cite{ferre2014sparklis} and SemFacet \cite{arenas2014semfacet}
that allow to explore the contents of semantic repositories through
a Faceted Search-like interface \cite{sacco2009dynamic,tzitzikas2016faceted}.
There are also approaches that translate free-text que\-ries to SPARQL
(like \cite{unger2012template}).
Providing such user-friendly interfaces on top of semantic layers
is out of the scope of this paper but
an important direction for future research.

\subsubsection{Analyzing Web Archives}

EverLast\cite{anand2009everlast}
is a web archiving framework built over a peer-to-peer architecture.
It supports human-assisted archive gathering and
allows for time-based search and analysis.
It indexes the documents by term and time
where each term is assigned to a peer responsible for managing its index.

Gossen et al. \cite{gossen2017extracting}
propose a method to extract interlinked event-centric document collections from large-scale web archives.
The proposed method relies on a specialized focused extraction algorithm
which takes into account both the temporal and the topical aspects of the documents.

Lin et. al. \cite{lin2014infrastructure} propose a platform
for analyzing web archives, called Warcbase,
which is built on Apache HBase\footnote{\url{https://hbase.apache.org/}},
a distributed data store.
Storing the data using HBase allows the use of tools in the
Hadoop ecosystem for efficient analytics and data processing.
Warcbase also provides  browsing capabilities similar to the Wayback Machine
allowing users to access historical versions of captured web pages.

Finally, ArchiveSpark \cite{holzmann2016archivespark} is a programming framework
for efficient and distributed web archive processing.
It is based on the Apache Spark
cluster computing framework\footnote{\url{https://spark.apache.org/}}
and makes use of standardized data formats
for analyzing web archives.
The \tool\ framework introduced in this paper is
an extension of ArchiveSpark for efficiently creating
semantic layers for web archives (more in Section \ref{subsec:framework}).

\section{Building Semantic Layers}
\label{sec:constrSemLayers}

\subsection{The \q{Open Web Archive} Data Model}
\label{subsec:semanticmodel}

\begin{figure*}
\centering \fbox{\includegraphics[width=5.7in]{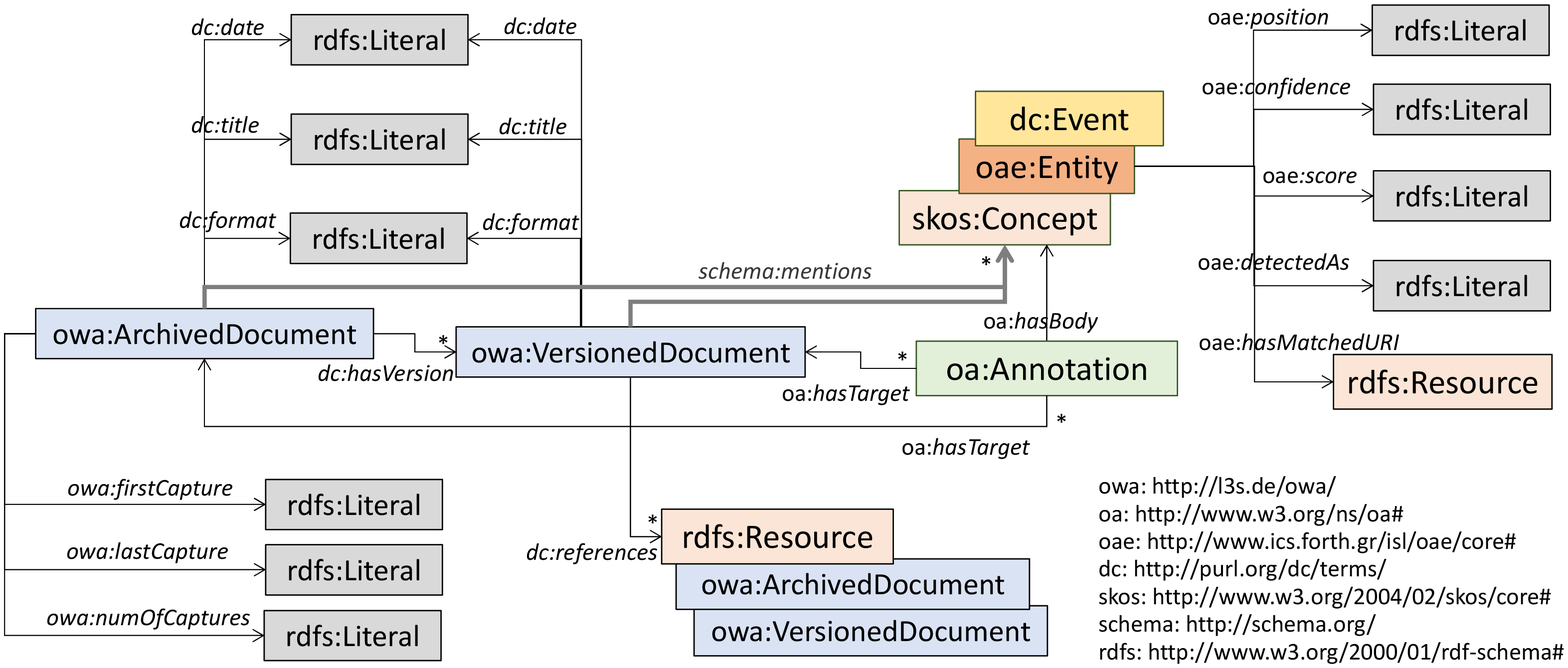}}
\vspace{-1mm} \caption{The {\em Open Web Archive} data model.}
\label{fig:owa}

\vspace{5mm}

\centering
\fbox{\includegraphics[width=5.7in]{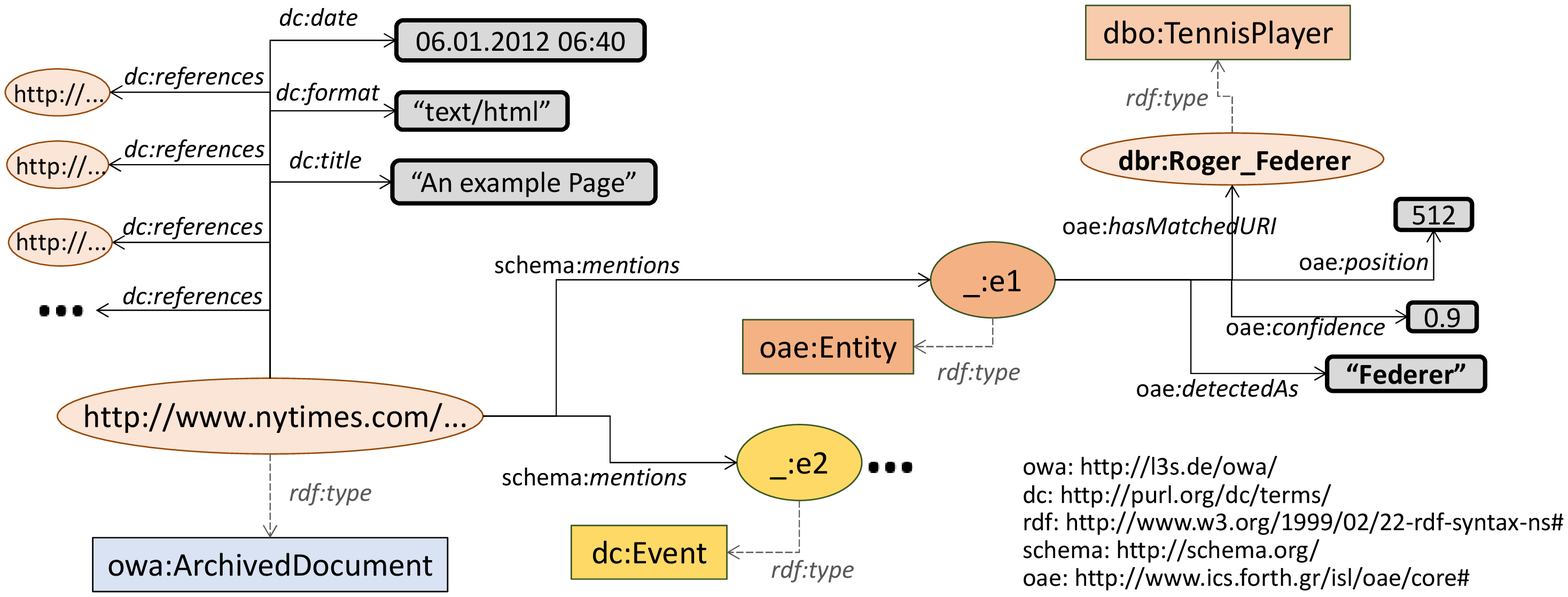}}
\vspace{-1mm} \caption{Describing an archived article
(non-versioned) using the {\em Open Web Archive} data model.}
\label{fig:owa_instNonVers}

\vspace{5mm}

\centering \fbox{\includegraphics[width=6.9in]{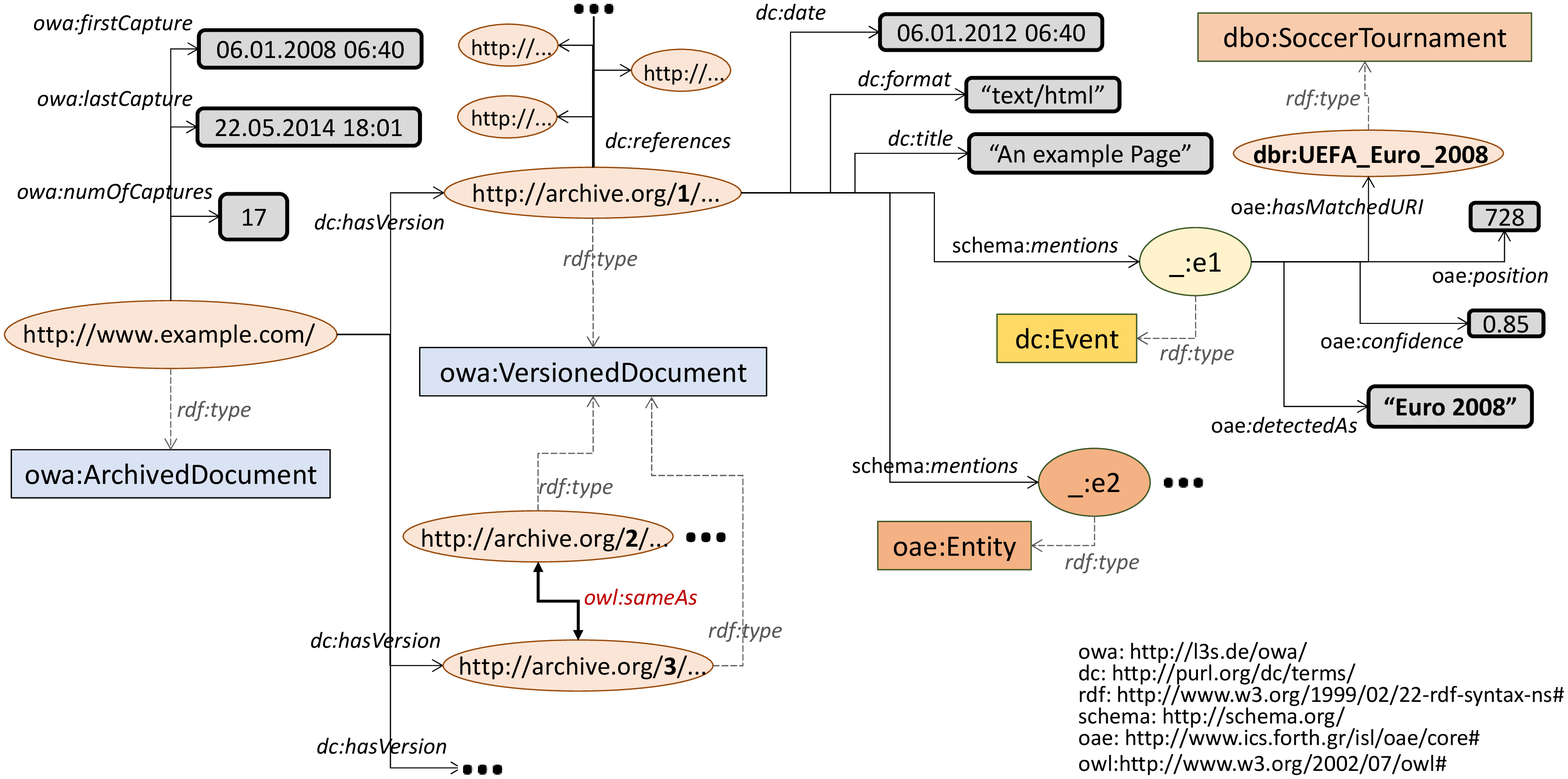}}
\vspace{-1mm} \caption{Describing an archived web page (versioned)
using the {\em Open Web Archive} data model.} \label{fig:owa_inst}
\end{figure*}

We first introduce an RDF/S data model
for describing metadata and semantic information about the documents of a web archive.
Figure \ref{fig:owa} depicts the proposed model,
which we call {\em Open Web Archive} data model.\footnote{The specification
 is available at: \url{http://l3s.de/owa/}}

We have defined 2 new classes and 3 new properties,
while we also exploit elements from other established data models.
The class {\tt owa:Archi\-ved\-Do\-cu\-ment} represents a document that has been archived.
An archived document may be linked or may not be linked with some versions,
i.e., instances of {\tt owa:Ve\-rsio\-ned\-Do\-cu\-ment}.
For example, an archi\-ved article from the
New York Times corpus \cite{sandhaus2008new} does not contain versions.
On the contrary,
Internet Archive
contains versions for billions of web sites.
For the case of versioned web archives,
and with correspondence to the Memento Framework (RFC 7089) \cite{van2013rfc},
an archived document actually corresponds to an {\em Original Resource}
and a versioned document to a {\em Memento}.
An archived document containing versions can be also associated with some metadata information
like the date of its first capture (using the property {\tt owa:firstCapture}),
the date of its last capture (using the property {\tt owa:lastCapture}) as well as
its total number of captures (using the property {\tt owa:numOfCaptures}).

An archived or versioned document
can be associated with three main kinds of elements:
i) with metadata information like date of publication or capture,
title of document, and format (mime type),
ii) with other archived or not documents (i.e., links to other web pages), and
iii) with a set of annotations.
For describing some of the metadata we exploit terms of the
Dublin Core Metadata Initiative\footnote{\url{http://dublincore.org/}}.
For describing an annotation, we exploit
the Open Annotation Data Model\footnote{\url{http://www.openannotation.org/spec/core/}} \cite{sanderson2013open}
and the Open Named Entity Extraction (NEE) Model\footnote{\url{http://www.ics.forth.gr/isl/oae/}} \cite{fafalios2015ijait}.
The Open Annotation Data Model specifies an RDF-based framework for creating associations (annotations)
between related resources, while the Open NEE Model is an extension
that allows describing the result of an entity extraction process.
An annotation has a {\em target}, which in our case is an archived or versioned document, and
a {\em body} which is an entity mentioned in the document.
We can also directly relate an archived or versioned document with an
entity by exploiting the property \q{{\em mentions}}
of schema.org\footnote{\url{http://schema.org/mentions}}.
This can highly reduce the number of derived triples.
An entity can be associated with information like
its name, a confidence score, its position in the document, and a resource (URI).
The URI enables to retrieve additional information from the Linked Open Data (LOD)
cloud \cite{heath2011linked} (like properties and relations with other entities).

Figure \ref{fig:owa_instNonVers} depicts an example of an archived non-versioned article.
We can see some of its metadata values (date, format, title),
its references to other web pages, and its annotations.
We notice that the entity name \q{Federer} was identified
in that document.
We can also see that this entity has been linked with the DBpedia resource
corresponding to the tennis player {\em Roger Federer}.
By accessing DBpedia, we can now retrieve more information about this entity
like its birth date, an image, a description in a specific language, etc.
Such links to DBpedia can also take the temporal aspect into account.
For example, we can provide entity URIs that lead to DBpedia entity descriptions
as they were at the time the web page was captured
(e.g., by exploiting DBpedia archives provided by Memento\footnote{\url{http://mementoweb.org/depot/native/dbpedia/}}).

Figure \ref{fig:owa_inst} depicts an example of an archived web page containing versions.
Now, each version has its own metadata, annotations and references to other web pages.
We notice that the event name \q{Euro 2008} was identified in the first version of the archived document
and was linked to the DBpedia resource corresponding to the soccer tournament {\em UEFA Euro 2008}.
The archived document is also associated with metadata information related to its versions.
Specifically we can see the date of its first capture, the date of its last capture and
its total number of captures.
In addition, by exploiting the {\em same-as} property of OWL Web ontology language \cite{bechhofer2009owl},
we can state that a specific version of a URL is the same as another version
(e.g., versions 2 and 3 in our example).
Thereby, we can avoid storing exactly the same data for two identical versions
(redundancy is a common problem in web archives).

\subsubsection*{Extensibility}
The proposed model is highly extensible.
For instance, we can
exploit the VoID Vocabulary \cite{alexander2009describing}
and express dataset-related information like
statistics (number of triples, number of entities, etc.),
creation or last modification date,
the subject of the dataset, and
collection from which the dataset was derived.
Likewise, one may exploit the PROV data model\footnote{\url{https://www.w3.org/TR/prov-dm/}}
and store provenance-related information
(e.g., which tool was used for crawling the documents or for annotating them,
what organizations or people were involved in the crawling or annotation process,
etc.).

\subsubsection*{Update}
Since the contents of the archived documents never change,
we can easily update a semantic layer by just
adding triples in the RDF repository.
For example, we can add triples that describe
more metadata about the archived documents,
or triples that describe more information about
the entities like properties, characteristics, or associations
with other entities.

For the case of versioned web archives,
we can also include new versions in the semantic layer.
However, in that case we should also
update the date of last capture
and the total number of captures of the corresponding
archived documents.

\subsection{The Construction Process}
\label{subsec:theprocess}

\begin{figure*}
\centering
\includegraphics[width=6in]{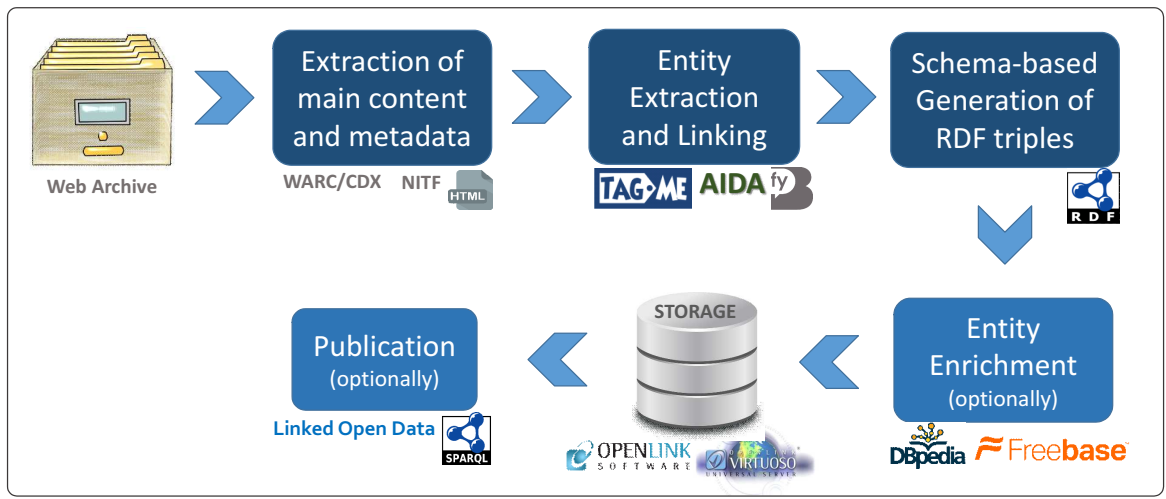}
\caption{The process of constructing a Semantic Layer.}
\label{fig:process}
\end{figure*}

Figure \ref{fig:process} depicts the
process of constructing a semantic layer.
The steps are the following:

\begin{itemize}
\item
{\em Reading of main content and metadata}.
We first extract the main content (full text) from
each archived document (for annotating it with entities) and
we also read its metadata.
This, of course, depends on the format used for
storing the archive. For example,
WARC (ISO 28500:2009)\footnote{\url{https://iipc.github.io/warc-specifications/specifications/warc-format/warc-1.0/}} is the standard format for storing web crawls,
CDX\footnote{\url{https://iipc.github.io/warc-specifications/specifications/cdx-format/cdx-2006/}} is widely used for describing metadata of web documents,
while NITF (News Industry Text Format)\footnote{\url{https://iptc.org/standards/nitf/}} is a standard
XML-based format for storing and sharing news articles.
For extracting the main content from HTML web pages,
we should also remove the surplus
around the main textual content of a web page (boilerplate, templates, etc.).
We can also extract any other information related to an archived document
that we may want to semantically describe, like the title of the web page or
links to other web pages.

\item
{\em Entity extraction and linking}.
We apply entity extraction and linking in the full text of each
archived document for detecting entities, events and concepts
mentioned in the document and associating them with web resources
(like DBpedia/Wikipedia URIs).
TagMe \cite{ferragina2010tagme},
AIDA \cite{hoffart2011robust} and
BabelFy \cite{moro2014entity} are
well-known entity extraction and linking tools with
 satisfactory performance in entity disambiguation.

\item
{\em Schema-based generation of RDF triples}.
Now, we exploit the {\em Open Web Archive} data model, as well
as any other needed vo\-ca\-bu\-la\-ry/onto\-lo\-gy, for
generating the RDF triples that
describe all the desired data related
to the archived documents (metadata, entities, etc.).
For representing the extracted entities (instances of {\tt oa:Annotation},
 {\tt oae:\-Entity}, {\tt dc:\-Event}, and {\tt dc:\-Concept}),
we can use blank nodes \cite{beckett2004rdf}
(since such information does not need to be assigned a unique URI).
We can use blank nodes for also naming the
archived or versioned documents (instances of {\tt owa:Archi\-ved\-Do\-cu\-ment} or
{\tt owa:\-Ve\-rsionedDo\-cu\-ment}) in case no URLs are given by the archive provider
and no other URLs can be used (e.g., links to the Wayback Machine).
Moreover,
for the case of versioned documents,
if a specific version of a document is the
same as an older version of the same document
(e.g., in case they have the same checksum),
we can add a {\em same-as} link
starting from the newer document and pointing to
the older one (thereby avoiding storing identical information).

\item
{\em Entity enrichment (optionally)}.
We can enrich the extracted entities with more
information coming from other knowledge bases
(like properties, characteristics and relations with other entities).
The LOD cloud contains hundreds of knowledge bases covering many domains.
In that way the semantic layer can directly offer more data about the extracted entities,
allowing for more sophisticated query capabilities
and faster query answering, without requiring
access to external knowledge bases.
This step can be also performed after the construction
of the semantic layer, at any time, since we
just have to add triples describing information about
the entities in the repository.

\item
{\em Storage}.
The derived RDF triples are stored in a triplestore
(e.g., OpenLink Virtuoso\footnote{https://virtuoso.openlinksw.com/}).
Now, we can access the triplestore and query the semantic layer
through SPARQL.

\item
{\em Publication (optionally)}.
We can make the triplestore publicly available through
a SPARQL endpoint and/or as Linked Data.
This will allow other applications to directly access and query
the semantic layer.
\end{itemize}

\subsection{The \q{ArchiveSpark2Triples} Framework}
\label{subsec:framework}

ArchiveSpark \cite{holzmann2016archivespark}
is a programming framework for efficiently analyzing
web archives stored in the standard WARC/ CDX format.
The core of ArchiveSpark is its unified data model
which stores records in an hierarchical way,
starting with the most essential metadata of a webpage like its URL, timestamp, etc.
Based on this metadata, ArchiveSpark can run basic operations
such as filtering, grouping and sorting very efficiently.
In a step-wise approach the records can be enriched
with more information by applying external modules,
called \textit{enrich functions}.
An \textit{enrich function} can call any third-party tool
to extract or generate new information from the contents of a web page.
These functions can be fully customized and shared among researchers and tasks.

\tool\footnote{\url{https://github.com/helgeho/ArchiveSpark2Triples}}
is an extension of ArchiveSpark that automates the
construction of a semantic layer.
It reads a web archive and outputs information about
its resources as well as derived information in the Notation3 (N3) RDF format
based on the {\em Open Web Archive} data model.
Internally, \tool\ defines three types of documents:
\textit{archived document} (instance of {\tt owa:Archi\-ved\-Do\-cu\-ment}),
\textit{versioned document} (instance of {\tt owa:Ve\-rsi\-oned\-Do\-cu\-ment}), and
\textit{same-as versioned document} (instance of {\tt owa:Ve\-rsi\-oned\-Do\-cu\-ment}
which constitutes a \textit{revisit-record}, i.e., duplicate of a previous capture).
In more detail:

\begin{itemize}
\item
An \textit{archived document} represents all versions of the same web page,
i.e., all records with the same URL.
Its triples reflect the web page as one unit,
including the number of captures in the web archive,
the timestamps of the first and last capture as well as pointers to
the corresponding \textit{versioned documents}.

\item
A \textit{versioned document} represents each individual capture of a web page,
i.e., every record of a web page in the archive.
The assignment of URLs to the versioned documents is customizable and thus can be defined by the user.
By default, the triples of such a document only include the date of the capture
and its mime type (e.g., text, image, etc.).
However, the framework supports to extend this easily
by accessing and transforming into triples any property of ArchiveSpark's data model.
If this step involves \textit{enrich functions},
the required content of the
web page is seamlessly integrated by ArchiveSpark's en\-ri\-ch\-ment me\-cha\-nisms.
In our case, we can use enrich functions to extract the title of a page, its
links to other pages, and its entities.
The extraction of entities requires an additional module
which uses the entity extraction and linking system Yahoo FEL \cite{BlancoWSDM2015}.
The corresponding enrich function is available
under {\em FEL4\-Archi\-ve\-Spark}\footnote{\url{https://github.com/helgeho/FEL4ArchiveSpark}}.

\item
A \textit{same-as versioned document} represents
an already archived web page whose content has not been changed.
In this case, a {\em same-as} property pointing to the previous record
is only created. The way in which duplicates are identified
is not part of the framework and can be defined as part of the generation workflow.
\end{itemize}

Finally, defining the vocabularies to use for producing the triples is part of the
generation workflow and thus can be customized by the user.
An example of a workflow is shown in Listing \ref{fig:workflowExample}\footnote{The
corresponding Jupyter Notebook is available at: \url{https://github.com/helgeho/ArchiveSpark2Triples/blob/master/notebooks/Triples.ipynb}}.

\renewcommand{\figurename}{Listing}
\setcounter{figure}{0}

\begin{figure*}[th]
\centering \scriptsize
\begin{Verbatim}[frame=lines,numbers=left,numbersep=1pt]
 import de.l3s.archivespark._
 import de.l3s.archivespark.implicits._
 import de.l3s.archivespark.specific.warc._
 import de.l3s.archivespark.specific.warc.specs._
 import de.l3s.archivespark.specific.warc.implicits._
 import de.l3s.archivespark.enrich._
 import de.l3s.archivespark.enrich.functions._
 import de.l3s.archivespark.enrich.dataloads._
 import de.l3s.archivespark.enrichfunctions.fel._
 import de.l3s.archivespark2triples._
 import org.apache.hadoop.io.compress.GzipCodec

 // Load the Entity Linking model (FEL)
 val modelFile = "english-nov15.hash"
 sc.setCheckpointDir("spark_checkpoint")
 sc.addFile("hdfs:///user/holzmann/" + modelFile)

 // Load the web archive collection (filter duplicates and very big records)
 ArchiveSpark2Triples.versionUrl = r => s"https://wayback.archive-it.org/2950/${r.timestamp}/${r.originalUrl}"
 val collection = "ArchiveIt-Collection-2950"
 val cdxPath = s"/data/archiveit/$collection/cdx/*.cdx.gz"
 val warcPath = s"/data/archiveit/$collection/warc"
 val raw = ArchiveSpark.load(sc, WarcCdxHdfsSpec(cdxPath, warcPath))
 val records = raw.distinctValue(_.get)((a, b) => a).filter(_.compressedSize < 1024 * 100).cache // 100 kb

 // Select successful responses of type HTML and detect duplicates
 val responses = records.filter(r => r.status == 200 && r.mime == "text/html")
 val earliestDigests = responses.map(r => ((r.surtUrl, r.digest), r)).reduceByKey{(r1, r2) => if (r1.time < r2.time) r1 else r2 }
 val duplicates = records.map(r => ((r.surtUrl, r.digest), r)).join(earliestDigests).map{case (_, records) => records}
        .filter{case (r1, r2) => r1.time != r2.time}

 // Generate ArchivedDocument triples representing distinct webpages
 val versions = earliestDigests.map{case (_, r) => r}.union(duplicates.map{case (r1, r2) => r1})
 val documentTriples = ArchiveSpark2Triples.generateDocs(versions)

 // Create "sameAs triples" from duplicates
 val sameAsTriples = ArchiveSpark2Triples.generateSameAsVersions(duplicates)

 // Generate VersionedDocument triples with title and entities
 val repartitioned = earliestDigests.map{case (_, r) => r}.repartition(5000)
 val title = HtmlText.of(Html.first("title"))
 val responsesWithTitles = repartitioned.enrich(title)
 val fel = FELwithTimeOut(scoreThreshold = -4, modelFile = modelFile).on(HtmlText)
 val responsesWithEntities = responsesWithTitles.enrich(fel)

 val versionTriples = ArchiveSpark2Triples.generateVersionsMapped(responsesWithEntities) {(record, uid, doc) =>
    val recordTitle = record.value(title).getOrElse("")
    val recordEntities = record.value(fel).getOrElse(Seq.empty)
    doc.appendTriples("dc:title", s"""\"$recordTitle\"""").appendChildren("schema:mentions", {
        recordEntities.zipWithIndex.map{case (entity, i) => TripleDoc(s"_:e$uid-$i", "oae:Entity", Seq(
            "oae:confidence" -> Seq(s""""${entity.score}"^^xsd:double"""),
            "oae:detectedAs" -> Seq(s"""\"${entity.span}\""""),
            "oae:position" -> Seq(s""""${entity.startOffset}"^^xsd:integer"""),
            "oae:hasMatchedURI" -> Seq(s"<http://dbpedia.org/resource/${entity.annotation}>")))}})}

 // Sort and store with headers
 val headers = TripleHeader.append("oae" -> "http://www.ics.forth.gr/isl/oae/core#")
 val triples = ArchiveSpark2Triples.toStringsSorted(headers, documentTriples, sameAsTriples, versionTriples)
 triples.saveAsTextFile(s"$collection-Triples1.gz", classOf[GzipCodec])
\end{Verbatim}
\vspace{-3.5mm}
\caption{An example of a workflow for generating a Semantic Layer.}
\label{fig:workflowExample}
\vspace{-2mm}
\end{figure*}

\subsubsection*{Efficiency}
\tool\ gains its efficiency from the efficiency of ArchiveSpark,
which is mainly a result of the two-way approach
that is used for data loading and access \cite{holzmann2016archivespark}.
An archived collection to be used with ArchiveSpark always consists of two parts,
the WARC files containing the data records with headers and payloads,
and the CDX files containing only basic metadata
such as URLs, timestamps and datatype (which are considerably smaller in size).
Hence, operations that rely exclusively on information contained in the metadata
can be performed very efficiently,
e.g., filtering out items of a certain type.
Eventually, if operations need to be performed on the actual contents,
only the required records are accessed using location pointers in the CDX files.
\tool\ benefits from this approach,
since records of a datatype other than {\em text/html},
such as images and videos, can be filtered out very fast.
In addition,
all properties of the {\em archived documents} and the majority of properties
of the {\em versioned documents} can be generated purely based on metadata and thus,
very efficiently.
In fact, the payload is accessed only for applying {\em enrich functions},
e.g., for extracting the title of a web page, its entities, etc.
However, these are only part of the {\em same-as versioned documents}
that do not constitute duplicates.

The most expensive task in our pipeline is the entity extraction process,
performed by {\em FEL4ArchiveSpark} using Yahoo FEL \cite{BlancoWSDM2015}
(a lightweight and very efficient entity extraction and linking system).
To avoid extraordinarily long runtimes,
{\em FEL4ArchiveSpark} supports to define a timeout
(set to 10 seconds per record in our experiments).
Additionally, we consider only web pages with a compressed size of less than 100 KB,
as larger file sizes are unlikely to constitute a web page and may indicate a malformed record.
Although the described steps are considered quite efficient,
the actual time for the entire workflow depends on the dataset size,
the nature of the data as well as the used computing infrastructure.
Indicatively, the Hadoop cluster used in our experiments for
producing a semantic layer for a web archive
of about 9 millions web pages
consisted of 25 compute nodes
with a total of 268 CPU cores and 2,688 GB RAM (more about this web archive in Section \ref{webArch}).
While the available resources strongly depend on
the load of the cluster and vary,
we worked with 110 executors in parallel most of the time,
which resulted in a runtime of 24 hours for
processing the entire collection of 474.6 GB of compressed WARC and CDX files.

\section{Case Studies}
\label{sec:casestudies}

In this section, we present three semantic layers
for three different types of web archives
and we showcase their query capabilities.
The semantic layers are publicly
available for experimentation and further research\footnote{\url{http://l3s.de/owa/semanticlayers/}}.

\subsection{A Semantic Layer for a News Archive}
\label{newsArch}
We created a semantic layer for the
New York Times (NYT) Annotated Corpus \cite{sandhaus2008new}
(a non-versioned news archive).
The corpus contains over 1.8 million articles published by
the NYT between 1987 and 2007.
We filtered out articles like memorial notices, corrections,
letters, captions, etc. which actually are not articles.
This reduced their number to 1,456,875.
For each article in the corpus,
a large amount of metadata is provided.
In this case study, we exploited only the article's {\em URL},
{\em title} and {\em publication date}.
Of course, one can exploit any other
of the provided metadata (like {\em author},
{\em taxonomic classifiers}, etc.) and extend the
semantic layer with more triples describing these metadata fields.

We used TagMe \cite{ferragina2010tagme} for extracting
entities from each article using a
confidence score of 0.2.
For each extracted entity, we stored its name (surface form),
its URI and its confidence score.
Table \ref{tbl:nyt-tagme} shows the number of articles and distinct entities per year.
In total, 856,283 distinct entities
of several types were extracted from the NYT articles.
Indicatively, the semantic layer has associated the articles with
304,502 distinct entities of type {\em Person} (i.e., of {\em rdf:type} \url{http://dbpedia.org/ontology/Person}),
86,237 of type {\em Location},
and 54,585 of type {\em Organization}.
Regarding the entities of type {\em Person},
63,537 are athletes, 23,974 are artists, and 8,818 are politicians.
The constructed semantic layer contains totally
195,958,390 triples.

\begin{table}[]
\centering
\caption{Number of articles and distinct entities per year contained in the Semantic Layer of the NYT corpus.}
\label{tbl:nyt-tagme}
\vspace{-2mm}
\setlength{\tabcolsep}{3pt}
\begin{tabular}{|c|c|c|c|}
\hline
\rowcolor[HTML]{EFEFEF}
{\color[HTML]{000000} Year} & {\color[HTML]{000000} Number of Articles} & {\color[HTML]{000000} Number of distinct Entities} \\ \hline
1987 & 98,311 & 201,245\\ \hline
1988 & 96,508 & 205,745\\ \hline
1989 & 94,465 & 201,362\\ \hline
1990 & 89,577 & 200,496\\ \hline
1991 & 74,298 & 188,402\\ \hline
1992 & 71,530 & 186,289\\ \hline
1993 & 67,320 & 185,423\\ \hline
1994 & 62,998 & 186,393\\ \hline
1995 & 71,944 & 188,716\\ \hline
1996 & 66,255 & 202,197\\ \hline
1997 & 57,395 & 199,935\\ \hline
1998 & 60,736 & 214,900\\ \hline
1999 & 61,014 & 218,546\\ \hline
2000 & 64,642 & 226,724\\ \hline
2001 & 66,838 & 219,858\\ \hline
2002 & 69,365 & 227,847\\ \hline
2003 & 66,833 & 226,896\\ \hline
2004 & 63,796 & 224,252\\ \hline
2005 & 62,822 & 228,426\\ \hline
2006 & 61,727 & 231,223\\ \hline
2007 & 28,501 & 151,119\\ \hline
\end{tabular}
\end{table}

\subsection{A Semantic Layer for a Web Archive}
\label{webArch}
Using \tool, we created a semantic layer
for the
\textit{Occupy Movement 2011/2012} collection\footnote{\url{https://archive-it.org/collections/2950}},
which has been generously provided to us by Archive-It.
The collection contains 9,094,573 captures of 3,036,326 web pages related to
protests and demonstrations around the world calling for social and economic equality.
For each version, we stored its {\em capture date},
its {\em title}, its {\em mime type} and its {\em extracted entities}
(using a confidence score of -4),
while for each distinct URL we stored its total number of captures,
the date of its first capture, and the date of its last capture.
For assigning URLs to the versioned web pages,
we used links to the collection's Wayback Machine provided by Archive-It.
In that way one can have direct online access to a specific version
of an archived web page.

The semantic layer contains 1,344,450 {\em same-as} properties,
which means that we avoided annotating and storing identical information
for a very large number of versioned web pages (about 15\% of all captures).
Moreover, 939,960 distinct entities (including concepts and events)
were extracted from the archived web pages.
For each entity, we stored its name (surface form),
its URI, its position in the text, and its confidence score.
The constructed semantic layer contains totally
more than 10 billion triples (10,884,509,868).

\subsection{A Semantic Layer for a Social Media Archive}
\label{tweetsArch}
We also created a semantic layer for a collection of tweets.
The collection comprises 1,363,415 tweets
posted in 2016 by 469 twitter accounts of USA newspapers.
For each tweet we exploit its {\em text}, {\em creation date},
{\em favorite count}, {\em retweet count}, and
the {\em screen name} of the account that posted the tweet.
For representing an instance of a tweet,
as well as its favorite and retweet count,
we used the OpenLink Twitter Ontology\footnote{\url{http://www.openlinksw.com/schemas/twitter}}
(its class {\tt Tweet} corresponds to an {\em archived document} in our model).

For extracting entities from the tweets,
we used Yahoo FEL (with confidence score -4).
For each extracted entity, we stored its name (surface form),
its URI and its confidence score.
In total, 146,854 distinct entities (including concepts and events)
were extracted from the collection.
Table \ref{tbl:twitter-fel} shows the number of tweets and distinct
entities per month.
The constructed semantic layer contains totally 19,242,761 triples.

\begin{table}[]
\centering
\caption{Number of tweets and distinct entities per month contained in the Semantic Layer of the Tweets Collection.}
\label{tbl:twitter-fel}
\vspace{-2mm}
\setlength{\tabcolsep}{3pt}
\begin{tabular}{|c|c|c|}
\hline
\rowcolor[HTML]{EFEFEF}
{\color[HTML]{000000} Month} & {\color[HTML]{000000} Number of Tweets} & {\color[HTML]{000000} Number of distinct Entities} \\ \hline
01.2016 & 110,250 & 56,244 \\ \hline
02.2016 & 111,370 & 57,557 \\ \hline
03.2016 & 123,445 & 61,184 \\ \hline
04.2016 & 114,697 & 59,844 \\ \hline
05.2016 & 112,273 & 59,476 \\ \hline
06.2016 & 109,412 & 58,043 \\ \hline
07.2016 & 113,105 & 55,248 \\ \hline
08.2016 & 115,255 & 57,998 \\ \hline
09.2016 & 116,391 & 59,251 \\ \hline
10.2016 & 118,429 & 59,179 \\ \hline
11.2016 & 114,651 & 56,103\\ \hline
12.2016 & 104,137 & 52,683\\ \hline
\end{tabular}
\end{table}

\subsection{Querying the Semantic Layers}
\label{queryCapabilities} By exploiting the expressive power of
SPARQL \cite{prud2008sparql} and its federated features
\cite{prud2013sparql,fafalios2016querying}, we can offer advanced
query capabilities over the semantic layers. Below we first discuss
how a semantic layer can satisfy the motivating questions described
in Section \ref{motivation} by also presenting interesting query
examples. We also present other exploitation scenarios for different
application contexts.

\subsubsection*{Information Exploration and Integration (Q1-Q2)}
A semantic layer allows running sophisticated que\-ries
that can also directly integrate information from external knowledge bases.
For example, Listing \ref{fig:queryExampleIntegrate} shows a SPARQL query
that can be answered by the semantic layer of the NYT corpus.
The query asks for articles of June 1989
discussing about New York lawyers born in Brooklyn.
By directly accessing DBpedia, the query retrieves the
entities that satisfy the query
as well as additional information (in our example the birth date
and a description in French of each lawyer).
The query returns  47 articles mentioning 5 different
New York lawyers born in Brooklyn.

\begin{figure}[th]
\centering \scriptsize
\begin{Verbatim}[frame=lines,numbers=left,numbersep=1pt]
SELECT ?article ?title ?date ?nylawyer ?bdate ?abstr WHERE {
 SERVICE <http://dbpedia.org/sparql> {
  ?nylawyer dc:subject dbc:New_York_lawyers ;
              dbo:birthPlace dbr:Brooklyn .
  OPTIONAL {
   ?nylawyer dbo:birthDate ?bdate ;
                dbo:abstract ?abstr FILTER(lang(?abstr)="fr")}}
 ?article dc:date ?date FILTER(?date>="1989-06-01"^^xsd:date
                                  && ?date<="1989-06-30"^^xsd:date)
 ?article schema:mentions ?entity .
 ?entity oae:hasMatchedURI  ?nylawyer .
 ?article dc:title ?title
} ORDER BY ?nylawyer
\end{Verbatim}
\vspace{-4mm}
\caption{SPARQL query for retrieving articles of June 1989
discussing about New York lawyers born in Brooklyn.}
\label{fig:queryExampleIntegrate}
\end{figure}

Listing \ref{fig:queryExampleTweet} shows a query that can be answered
by the semantic layer of the tweets collection.
The query requests the most popular tweets (having more than 50 retweets)
posted during the summer of 2016, mentioning
basketball players of the NBA team {\em Los Angeles Lakers}.
The query returns 14 tweets mentioning 7 different players.

\begin{figure}[th]
\centering \scriptsize
\begin{Verbatim}[frame=lines,numbers=left,numbersep=1pt]
SELECT DISTINCT ?tweet ?count ?date ?entityUri WHERE {
 SERVICE <http://dbpedia.org/sparql> {
   ?entityUri dc:subject dbc:Los_Angeles_Lakers_players }
 ?t a tw:Tweet ;
     dc:date ?date FILTER(?date>="2016-06-01"^^xsd:dateTime &&
                               ?date<="2016-08-31"^^xsd:dateTime)
 ?t tw:retweetCount ?count FILTER (?count > 50) .
 ?t schema:text ?tweet ; schema:mentions ?entity .
 ?entity oae:hasMatchedURI ?entityUri }
 \end{Verbatim}
\vspace{-4mm}
\caption{SPARQL query for retrieving popular tweets of summer 2016
mentioning basketball players of Los Angeles Lakers.}
\label{fig:queryExampleTweet}
\end{figure}

We can also combine information coming from
different semantic layers.
For example, the query in Listing \ref{fig:queryCombine} requests
tweets of summer 2016 mentioning
basketball players of Los Angeles Lakers discussed in articles of the
same time period.

\begin{figure}[th]
\centering \scriptsize
\begin{Verbatim}[frame=lines,numbers=left,numbersep=1pt]
SELECT DISTINCT ?player ?tweet WHERE {
 SERVICE <http://dbpedia.org/sparql> {
   ?player dc:subject dbc:Los_Angeles_Lakers_players }
 ?article dc:date ?date FILTER(?date>="2016-06-01"^^xsd:date
                                  && ?date<="2016-08-31"^^xsd:date)
 ?article schema:mentions ?articleEntity .
 ?articleEntity oae:hasMatchedURI ?player .
 ?tweet a tw:Tweet ;
        dc:date ?date FILTER(?date>="2016-06-01"^^xsd:date
                              && ?date<="2016-08-31"^^xsd:date) .
 ?tweet schema:mentions ?tweetEntity .
 ?tweetEntity oae:hasMatchedURI ?player }
\end{Verbatim}
\vspace{-4mm}
\caption{SPARQL query for retrieving tweets of summer 2016
mentioning players of Los Angeles Lakers discussed in articles of the same time period.}
\label{fig:queryCombine}
\end{figure}

\subsubsection*{Information Inference (Q3)}
By querying a semantic layer we can infer
useful knowledge related to the archived documents
that is very laborious to derive otherwise.
For example,
Listing \ref{fig:queryExampleInfer} shows a query that can be answered by the semantic layer
of the {\em Occupy Movement} collection.
The query asks for the most discussed journalists in the
web pages of this collection.
Notice that the query counts the archived documents, not the versions.
In that way we avoid counting multiple times exactly the same pages
captured in different time periods.
The query returns {\em Ralph Nader},
{\em Chris Hedges} and {\em Dylan Ratigan},
as three of the most discussed journalists.

\begin{figure}[th]
\centering \scriptsize
\begin{Verbatim}[frame=lines,numbers=left,numbersep=1pt]
SELECT ?journ (COUNT(DISTINCT ?page) AS ?num) WHERE {
 SERVICE <http://dbpedia.org/sparql> {
   ?journ a yago:Journalist110224578 }
 ?page a owa:ArchivedDocument ;
           dc:hasVersion ?version .
 ?version schema:mentions ?entity .
 ?entity oae:hasMatchedURI  ?journ .
} GROUP BY ?journ ORDER BY DESC(?num)
\end{Verbatim}
\vspace{-4mm}
\caption{SPARQL query for retrieving the most discussed
journalists in web pages of the {\em Occupy Movement} collection.}
\label{fig:queryExampleInfer}
\end{figure}

Likewise, by running a query at the semantic layer of the NYT corpus
requesting the number of articles per year discussing
about {\em Nelson Mandela} (Listing \ref{fig:queryExampleInferMand}),
we can see that in 1990 the number of articles is much higher compared
to the previous years, meaning that this year was probably important
for {\em Nelson Mandela} (indeed, as in
1990 {\em Nelson Mandela} was released from prison).

\begin{figure}[th]
\centering \scriptsize
\begin{Verbatim}[frame=lines,numbers=left,numbersep=1pt]
SELECT ?year (COUNT(DISTINCT ?article) AS ?num) WHERE {
  ?article dc:date ?date ;
             schema:mentions ?entity .
  ?entity oae:hasMatchedURI dbr:Nelson_Mandela
} GROUP BY (year(?date) AS ?year) order by ?year
\end{Verbatim}
\vspace{-4mm}
\caption{SPARQL query for retrieving the number of articles per
year mentioning {\em Nelson Mandela}.}
\label{fig:queryExampleInferMand}
\end{figure}

Listing \ref{fig:queryExampleDrugs} shows another example in which
the query requests the most discussed drugs in articles of 1987.
The query returns the following top-5 drugs:
Cocaine (778 articles),
Heroin (248 articles),
Aspirin (63 articles),
Zidovudine (53 articles),
Furosemide (53 articles).
If we run the same query for the year 1997, the results are:
Cocaine (462 articles),
Heroin (275 articles),
Nicotine (125 articles),
Fluoxetine (61 articles),
Caffeine (58 articles).
We notice that Cocaine and Heroin remain the two most discussed drugs,
however we also see that Nicotine is highly discussed in 1997 but not in 1987.

\begin{figure}[th]
\centering \scriptsize
\begin{Verbatim}[frame=lines,numbers=left,numbersep=1pt]
SELECT ?drug (count(DISTINCT ?article) as ?numOfArticles) WHERE {
  SERVICE <http://dbpedia.org/sparql> {
    ?drug a dbo:Drug }
  ?article dc:date ?date FILTER(year(?date) = "1987") .
  ?article schema:mentions ?ent .
  ?ent oae:hasMatchedURI  ?drug .
} GROUP BY ?drug ORDER BY DESC(?numOfArticles)
\end{Verbatim}
\vspace{-4mm}
\caption{SPARQL query for retrieving the most discussed drugs in 1987.}
\label{fig:queryExampleDrugs}
\end{figure}

\subsubsection*{Robustness and Multilinguality (Q4-Q5)}
Each entity extracted from the archived documents
is assigned a unique URI (together with a confidence score)
which can be used for retrieving documents and information related to that entity.
This means that all different mentions of an entity
(e.g., name variants or names in different languages) are assigned the
same unique URI.
Thereby, we can query a semantic layer and retrieve information
related to one or more entities without having to worry
about the names of the entities (like in the queries
of Listings \ref{fig:queryExampleIntegrate}-\ref{fig:queryExampleDrugs}).
Of course, this also depends on the entity linking
system used for extracting the entities, specifically
on its \q{time-awareness} and correct disambiguation
(e.g., for understanding that {\em Leningrad} corresponds to the DBpedia URI
\url{http://dbpedia.org/resource/Saint_Petersburg}), as well as on
whether it supports
the identification of entities in different languages
(e.g., for assigning the same URI \url{http://dbpedia.org/resource/Abortion} to
both \q{abortion} and \q{Sch\-wa\-nger\-schaft\-sabbruch}).

\subsubsection*{Interoperability (Q6)}
RDF is a standard model for data interchange on the Web
and has features that facilitate data integration.
Describing metadata and content-based information about web archives in RDF
makes their contents machine {\em understandable}, and
allows their direct exploitation by other systems and tools.
Moreover, following the LOD principles for publishing a semantic layer
enables other systems to directly access it,
while the advanced query capabilities that it offers
allow the easy identification of an interesting part of a web archive
(related to a time period and some entities)
by just writing and submitting a SPARQL query.

\subsubsection*{Other exploitation scenarios}

{\em Time-Aware Entity Recommendation.}
Recent works have shown that entity recommendation is time-dependent,
while the co-occurrence of entities in documents of a given time period is
a strong indicator of their relatedness during that period and
thus should be taken into consideration \cite{zhang2016probabilistic,tran2017beyond}.
By querying a semantic layer,
we can easily find entities of a specific type, or having some specific characteristics,
that co-occur frequently with a given entity in a specific time period,
thereby enabling the provision of time and context aware entity recommendations.
For example, the query in Listing \ref{fig:queryCooccured} retrieves the top-5 politicians
co-occurring with {\em Barack Obama} in NYT articles of summer 2007.
Here one could also apply a more sophisticated approach, e.g., by also considering
the inverse document frequency of the co-occurred entities in the same time-period.

\begin{figure}[th]
\centering \scriptsize
\begin{Verbatim}[frame=lines,numbers=left,numbersep=1pt]
SELECT ?politician (count(distinct ?article) as ?num) WHERE {
  SERVICE <http://dbpedia.org/sparql> {
     ?politician a dbo:Politician }
  ?article dc:date ?date FILTER(?date >= "2007-06-01"^^xsd:date &&
                                       ?date <= "2007-08-30"^^xsd:date)  .
  ?article schema:mentions ?entity .
  ?entity oae:hasMatchedURI dbr:Barack_Obama .
  ?article schema:mentions ?entityPolit.
  ?entityPolit oae:hasMatchedURI ?politician
                           FILTER (?politician != dbr:Barack_Obama)
} GROUP BY ?politician ORDER BY DESC(?num) LIMIT 5
\end{Verbatim}
\vspace{-4mm}
\caption{SPARQL query for retrieving the top-5 politicians
co-occurring with {\em Barack Obama} in NYT articles of summer 2007.}
\label{fig:queryCooccured}
\end{figure}

\vspace{2mm} \noindent
{\em Evolution of entity-related features.}
The work in \cite{fafalios2017tpdl} has proposed a set of measures that
allow studying how entities are reflected in a social media archive and how
entity-related information evolves over time.
Given an entity and a time period, the proposed measures capture the following entity aspects:
{\em popularity},
{\em attitude} (predominant sentiment),
{\em sentimentality} (magnitude of sentiment),
{\em controversiality}, and
{\em connectedness} to other entities.
Such time-series data can be easily computed by running SPARQL queries
on a corresponding semantic layer (considering also that the layer contains the sentiments of the tweets).
For example, the query in Listing \ref{fig:obamaPopularity} retrieves
the monthly popularity  of {\em Barack Obama} in tweets of 2016 (using Formula 1 of \cite{fafalios2017tpdl}).

\begin{figure}[th]
\centering
\scriptsize
\begin{Verbatim}[frame=lines,numbers=left,numbersep=1pt]
SELECT ?month xsd:double(?cEnt)/xsd:double(?cAll)
WHERE {
 { SELECT (month(?date) AS ?month) (count(?tweet) AS ?cAll) WHERE {
     ?tweet dc:date ?date FILTER(year(?date) = 2016)
   } GROUP BY month(?date) }
 { SELECT (month(?date) AS ?month) (count(?tweet) AS ?cEnt) WHERE {
     ?tweet dc:date ?date FILTER(year(?date) = 2016) .
     ?tweet schema:mentions ?entity .
     ?entity oae:hasMatchedURI dbr:Barack_Obama
   } GROUP BY month(?date) }
} ORDER BY ?month
\end{Verbatim}
\vspace{-4mm}
\caption{SPARQL query for retrieving the monthly popularity of {\em Barack Obama} in tweets of 2016.}
\label{fig:obamaPopularity}
\end{figure}

\vspace{2mm} \noindent
{\em Identification of Similar or Identical Documents.}
We can find similar documents by comparing the entities mentioned on them.
The idea is that if two documents mention a big number of common entities
then they are probably about the same topic.
For example, given a NYT article about golf,
the query in Listing \ref{fig:similarDocs} retrieves the top-5 documents with the
bigger number of common entities.
By inspecting the returned results, we notice that all are about golf.

\begin{figure}[th]
\centering
\scriptsize
\begin{Verbatim}[frame=lines,numbers=left,numbersep=1pt]
SELECT ?article2 (count(?entUri2) as ?numOfCommon) WHERE {
 nyt:9504E4D71530F932A35755C0A9619C8B63 schema:mentions ?entity1 .
 ?entity1 oae:hasMatchedURI ?entUri1 .
 ?article2 schema:mentions ?entity2
      FILTER (?article2 != nyt:9504E4D71530F932A35755C0A9619C8B63)
 ?entity2 oae:hasMatchedURI ?entUri2 FILTER(?entUri2 = ?entUri1) .
} GROUP BY ?article2 ORDER BY DESC(?numOfCommon) LIMIT 5
\end{Verbatim}
\vspace{-4mm}
\caption{SPARQL query for retrieving similar documents.}
\label{fig:similarDocs}
\end{figure}

Likewise, we can find possibly identical documents by checking if they
contain exactly the same number of occurrences of the same entities.
This can be especially useful for the case of versioned web archives where
two versions of the same web page may have the same main content
but different checksums because, for example, of different layout.

\vspace{2mm} \noindent
{\em Advancing Information Retrieval.}
Recent works have shown that the exploitation of
entities extracted from search results can enhance the effectiveness of keyword-based
search systems in different contexts,
like in biomedical \cite{fafalios2017jasist} and academic \cite{xiong2017explicit} search.
Consequently, a semantic layer built on top of a collection of archived documents
can also serve a search system operating over the same collection.

\section{Evaluation}
\label{sec:eval}

Our objective is to show that
for a bit more complex information needs (e.g., of exploratory nature),
keyword-based search systems return poor results and thus
there is the need for more advanced information seeking strategies.
This corresponds to our first motivating question ({\bf Q1}).
We also study the quality of the results returned by a semantic layer
(for identifying possible problems and limitations)
as well as the efficiency of query answering.

\begin{table*}[]
\centering
\caption{List of information needs and corresponding free-text queries used in the evaluation.}
\vspace{-1mm}
\setlength{\tabcolsep}{2.5pt}
\label{tbl:info-needs}
\begin{tabular}{|c|p{120mm}|p{44mm}|}
\hline
\rowcolor[HTML]{EFEFEF}
\textbf{\#}              & \textbf{Information Need}                                                                                 & \textbf{Free-text query}                     \\ \hline
1                        & Find articles of June-August 1998 mentioning actors winners of an academy award for Best Actor           & Best actor academy award winner              \\ \hline
2                        & Find articles of July-August 1989 mentioning players of Los Angeles Lakers (NBA team)                    & Los Angeles Lakers player                    \\ \hline
3                        & Find articles of August 1992 mentioning African-American film producers                            & African-American film producer               \\ \hline
4                        & Find articles of 5-8/1/1990 mentioning drugs which act as stimulants                                     & Stimulant drugs                              \\ \hline
5                        & Find articles of 1/7/1992-20/9/1992 mentioning Ferrari Formula One drivers                               & Ferrari formula one drivers                  \\ \hline
6                        & Find articles of 5/7/1989-15/8/1989 mentioning assassinated Indian politicians                           & Assassinated Indian politicians              \\ \hline
7                        & Find articles of 1-19/06/1990 mentioning American crime thriller films                                   & American crime thriller films                \\ \hline
8                        & Find articles of July-August 1989 mentioning Boing 747 aircraft accidents                                & boeing 747 aircraft accidents                \\ \hline
9                        & Find articles of 1/7/1994-18/9/1994 mentioning Australian cricketers who played One Day Internationals   & Australian cricketers one day internationals \\ \hline
10                       & Find articles of 4/7/1995 mentioning companies listed on the New York Stock Exchange (NYSE)              & Companies listed on NYSE                     \\ \hline
11                       & Find articles of 1/7/1994-15/8/1994 mentioning video-game consoles                                       & Video game consoles                          \\ \hline
12                       & Find articles of 1/7/1992-15/9/1992 mentioning famous Indian personalities who received Padma Shri Award & Indian Padma Shri recipients                 \\ \hline
\multicolumn{1}{|l|}{13} & Find articles of July-September 1993 mentioning bacterial sexually transmitted diseases                  & Bacterial stds                               \\ \hline
\multicolumn{1}{|l|}{14} & Find articles of July 1989 mentioning operations of the Central Intelligence Agency (CIA)                & CIA operations                               \\ \hline
\multicolumn{1}{|l|}{15} & Find articles of 1/8/1998 mentioning Grammy Award Winners                                               & grammy award winner                          \\ \hline
\multicolumn{1}{|l|}{16} & Find articles of 1989 mentioning Indian meat dishes                                                      & Indian meat dishes                           \\ \hline
\multicolumn{1}{|l|}{17} & Find articles of July-September 1994 mentioning mammalian animals found in India                         & Indian mammals                               \\ \hline
\multicolumn{1}{|l|}{18} & Find articles of 1-10/7/1989 mentioning US fast food chains                                              & US fast food chains                          \\ \hline
\multicolumn{1}{|l|}{19} & Find articles of 1/7/1997-2/8/1997 mentioning NASA civilian astronauts                                   & NASA civilian astronauts                     \\ \hline
\multicolumn{1}{|l|}{20} & Find articles of 1/07/1989-15/8/1989 mentioning geological hazards                                       & geological hazard                            \\ \hline
\end{tabular}
\end{table*}

\subsection{Setup}
We have defined a set of 20 information needs of
exploratory nature.
Each information need requests documents
of a specific time period, related to some entities of interest.
We used the NYT corpus as the underlying archived collection.
For example, {\em \q{find articles of August 1992 mentioning African-American film producers}}
is such an {\em exploratory} information need.

Each of the information needs corresponds to a SPARQL query
and to a free-text query that better describes the information need
(in our evaluation we consider one interaction step,
i.e., one submitted query).
As an example, for the information need
{\em \q{find articles of August 1992 discussing about African-American film producers}},
the free-text query that is used is {\em \q{African-American film producer}}
(we manually specify the date range to each system).
Table \ref{tbl:info-needs} shows the full list of information needs and the corresponding free-text queries.

We evaluated and compared the results returned by the SPARQL query over the semantic layer
with the results returned by the following two keyword-based search systems operating over the NYT corpus:
a) Google News (adding at the end of the query the string \q{site:nytimes.com} for
returning only results from this domain),
b) HistDiv \cite{singh2016history}, which uses a different, diversity-oriented approach
for searching news archives.
Moreover, in the reported results we did not
consider 23 articles (out of totally 356 articles)
returned by the SPARQL queries because they do not
exist in Google News.

For each information need, we measure:
\begin{itemize}
\item  the number of hits returned by the SPARQL query
\item  the number of {\em relevant} hits returned by the SPARQL query
\item  the number of hits returned by each search system
\item  the number of {\em relevant} hits returned by each search
        system, existing in the set of relevant hits returned by the SPARQL query
\item The number of {\em relevant} hits returned by each search system, \underline{not}
existing in the set of relevant hits returned by the SPARQL query.
\end{itemize}
The SPARQL queries that correspond to the 20 information needs
as well as the full results and the relevance judgements are publicly
available\footnote{\url{http://l3s.de/owa/semanticlayers/SemLayerEval.zip}}.

\begin{table*}[]
\centering
\small
\caption{Comparative evaluation results on effectiveness.}
\vspace{-1mm}
\setlength{\tabcolsep}{3pt}
\label{tbl:eval}
\begin{tabular}{l|l|r|r|r|r|r|r|r|r|r|r|r|r|r|r|r|r|r|r|r|r|}
\hline
\multicolumn{2}{|r|}{\cellcolor[HTML]{EFEFEF}\textbf{Query:}}                                                                                                                          & \cellcolor[HTML]{EFEFEF}\textbf{1} & \cellcolor[HTML]{EFEFEF}\textbf{2} & \cellcolor[HTML]{EFEFEF}\textbf{3} & \cellcolor[HTML]{EFEFEF}\textbf{4} & \cellcolor[HTML]{EFEFEF}\textbf{5} & \cellcolor[HTML]{EFEFEF}\textbf{6} & \cellcolor[HTML]{EFEFEF}\textbf{7} & \cellcolor[HTML]{EFEFEF}\textbf{8} & \cellcolor[HTML]{EFEFEF}\textbf{9} & \cellcolor[HTML]{EFEFEF}\textbf{10} & \cellcolor[HTML]{EFEFEF}\textbf{11} & \cellcolor[HTML]{EFEFEF}\textbf{12} & \cellcolor[HTML]{EFEFEF}\textbf{13} & \cellcolor[HTML]{EFEFEF}\textbf{14} & \cellcolor[HTML]{EFEFEF}\textbf{15} & \cellcolor[HTML]{EFEFEF}\textbf{16} & \cellcolor[HTML]{EFEFEF}\textbf{17} & \cellcolor[HTML]{EFEFEF}\textbf{18} & \cellcolor[HTML]{EFEFEF}\textbf{19} & \cellcolor[HTML]{EFEFEF}\textbf{20} \\ \hline \hline
\multicolumn{1}{|c|}{\multirow{2}{*}{SPARQL}}                                                & \begin{tabular}[c]{@{}l@{}}Num of results\end{tabular}                                     & 27 & 34 & 37 & 16 & 11 & 14 & 18 & 8 & 11 & 15 & 15 & 12 & 13 & 16 & 14 & 12 & 15 & 13 & 16 & 15 \\ \cline{2-22}
\multicolumn{1}{|c|}{}                                                                       & \begin{tabular}[c]{@{}l@{}}Num of {\bf relevant} results\end{tabular}                            & 27 & 29 & 35 & 16 & 9  & 14 & 4  & 8 & 1  & 15 & 2  & 8  & 13 & 16 & 13  & 10 & 15 & 13 & 15 & 15 \\ \hline \hline
\multicolumn{1}{|c|}{\multirow{3}{*}{\begin{tabular}[c]{@{}l@{}}Google\\News\end{tabular}}}   & \begin{tabular}[c]{@{}l@{}}Num of results\end{tabular}                                     & 8  & 1  & 0  & 0  & 0  & 1  & 1  & 1 & 0  & 0  & 0  & 0  & 0  & 2  & 0  & 6  & 1  & 1  & 1  & 1  \\ \cline{2-22}
\multicolumn{1}{|l|}{}                                                                       & \begin{tabular}[c]{@{}l@{}}Num of {\bf relevant} results\\returned by SPARQL\end{tabular}     & 0  & 0  & 0  & 0  & 0  & 0  & 0  & 0 & 0  & 0  & 0  & 0  & 0  & 0  & 0  & 0  & 1  & 0  & 0  & 0  \\ \cline{2-22}
\multicolumn{1}{|l|}{}                                                                       & \begin{tabular}[c]{@{}l@{}}Num of {\bf relevant} results\\{\bf not} returned by SPARQL\end{tabular} & 0  & 1  & 0  & 0  & 0  & 0  & 0  & 0 & 0  & 0  & 0  & 0  & 0  & 0  & 0  & 0  & 0  & 0  & 0  & 1  \\ \hline \hline
\multicolumn{1}{|c|}{\multirow{3}{*}{HistDiv}}                                               & \begin{tabular}[c]{@{}l@{}}Num of results\end{tabular}                                     & 0  & 3  & 1  & 0  & 0  & 0  & 0  & 4 & 0  & 0  & 0  & 0  & 0  & 0  & 0  & 25 & 2  & 0  & 0  & 0  \\ \cline{2-22}
\multicolumn{1}{|l|}{}                                                                       & \begin{tabular}[c]{@{}l@{}}Num of {\bf relevant} results\\returned by SPARQL\end{tabular}     & 0  & 2  & 0  & 0  & 0  & 0  & 0  & 0 & 0  & 0  & 0  & 0  & 0  & 0  & 0  & 3  & 1  & 0  & 0  & 0  \\ \cline{2-22}
\multicolumn{1}{|l|}{}                                                                       & \begin{tabular}[c]{@{}l@{}}Num of {\bf relevant} results\\{\bf not} returned by SPARQL\end{tabular} & 0  & 1  & 0  & 0  & 0  & 0  & 0  & 3 & 0  & 0  & 0  & 0  & 0  & 0  & 0  & 3  & 0  & 0  & 0  & 0  \\ \hline
\end{tabular}
\end{table*}

\subsection{Results}
Table \ref{tbl:eval} shows the results.
We notice that the keyword-based search systems
cannot retrieve many relevant hits, while for many cases
the number of returned results is zero.
This illustrates that their effectiveness is
poor for more advanced information needs like those in our experiments
(considering however that we allow one interaction step).
The reason for this poor performance is the fact that each information
need describes a category of entities which refers to a number of
(possibly unknown) entities,
while the corresponding free-text query does not contain the entity names.
For example, the query {\em \q{African-American film producer}}
does not contain the actual names of any of these film producers.
Note that during an exploratory search process,
users may be unfamiliar with the domain of their goal
(e.g., they may not know the names of the entities of interest),
may be unsure about the ways to achieve their goal (e.g., not sure about the query to submit
to a search system),
or may need to learn about the topic in order to understand how to achieve their
goal (e.g., learn facts about some entities of interest) \cite{marchionini2006exploratory}.
For achieving a better performance, the user should probably
first find entities belonging to the corresponding information need and
then submit many queries using the entity names in the query terms.
Thus, multiple interaction and exploration steps may be needed.
However this can be infeasible, for example in case
of a large number of entities of interest.

Nevertheless, the results also show that in a few cases
the search system returns relevant hits that
are not returned by the SPARQL query
(e.g., \#2 and \#20 for Google, \#2, \#8 and \#16 for HistDiv).
In addition, some of the hits returned by the SPARQL query
are not relevant (e.g., 5 results of \#2),
while especially in three cases (\#7, \#9, and \#11),
this number is very large.
This is due to disambiguation error
of the entity linking system.
For example, for the information need \#9
({\em \q{Find articles discussing about Australian Cricketers
who played One Day Internationals}}),
the entity extraction system wrongly linked the name \q{John Dyson}
to the former international cricketer John Dyson,
instead of the deputy mayor John Dyson (at the time of Rudolph Giuliani's mayoralty)
discussed in the articles.
Therefore, the performance of the entity extraction system as well as
the confidence threshold used for entity disambiguation can
affect the quality  of the retrieved results.
Applying a low confidence threshold can increase recall, however
many irrelevant hits may also be returned.
On the contrary, by applying a high confidence threshold,
the returned results are less but the probability that they are correct is higher.

Table \ref{tbl:failure} details all the failure cases.
In summary, we have identified the following problems
that can affect the quality of the results:

\begin{itemize}
\item
{\em False positive:} A SPARQL query may return a
result which is not relevant, due to disambiguation error
of the underlying entity linking system.

\item
{\em False negative:} A SPARQL query may not return a relevant result because:
i) the entity linking system did not manage to recognize one of the entities of interest,
ii) the entity linking system did not disambiguate correctly
an extracted entity of interest,
iii) the confidence score of the extracted entity of interest
is under the threshold used for entity disambiguation.

\item
{\em Temporal inconsistency:}
A SPARQL query may return an irrelevant hit
or may not return a relevant hit,
because a property of an entity of interest has changed value.
For example,
the query of Listing \ref{fig:queryExampleTweet}
may return a tweet for a basketball player who was playing
in a different team at the time the tweet was posted
(although this also depends on user's intention, since
he/she may be interested in also such players).
Likewise, a query may not return a hit because
the knowledge base (from which we retrieve the list of players)
may not contain information about the team's old players.
Thus, the contents of the knowledge base, its \q{freshness}
and its completeness, affect the quality
of the retrieved results.
\end{itemize}

\begin{table*}[]
\centering
\caption{Detailed analysis of SPARQL failure cases.}
\vspace{-1mm}
\setlength{\tabcolsep}{4pt}
\label{tbl:failure}
\begin{tabular}{|c|p{155mm}|}
\hline
\rowcolor[HTML]{EFEFEF}
\multicolumn{1}{|l|}{\cellcolor[HTML]{EFEFEF}\textbf{Query}} & \textbf{SPARQL Failure Analysis}                                                                                                                                                                                                                                                                                                                                                    \\ \hline
2                                                            & SPARQL returns 5 irrelevant results (disambiguation errors of the mentions {\em \lq{}Malcolm C.', \lq{}Leonard C. Green', \lq{}Jon Barry', \lq{}Bobby Duhon', \lq{}Kevin McKenna'}). Google and HistDiv return 1 relevant result which though is not returned by SPARQL. The result mentions {\em \lq{}Mike Johnson'} which is not linked by the entity linking system. \\ \hline
3                                                            & SPARQL returns 2 irrelevant results (disambiguation errors of the mentions {\em \lq{}Michael Jackson'} and {\em \lq{}ice cube'}).                                                                                                                                                                                                                                                                       \\ \hline
5                                                            & SPARQL returns 2 irrelevant results (disambiguation errors of the mentions {\em \lq{}Tony Brooks'} and {\em \lq{}Pedro Rodriquez'}).                                                                                                                                                                                                                                                                    \\ \hline
7                                                            & SPARQL returns 14 irrelevant results (disambiguation errors of the mentions {\em \lq{}avenging angel', \lq{}Thomas King of New York', \lq{}don't say a word', \lq{}man apart', \lq{}usual suspects', \lq{}running scared', \lq{}training day', \lq{}ten to midnight', \lq{}Black Dahlia'}).                                                                                                                                   \\ \hline
8                                                            & HistDiv returns 3 relevant results which are not returned by SPARQL. The results mention {\em Boeing 747} accidents which though are not linked by the entity linking system.                                                                                                                                                        \\ \hline
9                                                            & SPARQL returns 10 irrelevant results (disambiguation errors of the mentions {\em \lq{}Kevin Wright', \lq{}Alan Conolly', \lq{}Matthew Elliott', \lq{}John Dyson'}).                                                                                                                                                                                                                                   \\ \hline
11                                                           & SPARQL returns 13 irrelevant results (disambiguation errors of the mentions {\em \lq{}a can', \lq{}3DO', \lq{}NES', \lq{}Wii'}).                                                                                                                                                                                                                                                                           \\ \hline
12                                                           & This is a special case: 4 results mention the actor Ben Kingsley. Ben Kingsley is from England however he has been awarded the Padma Shri award.                                                                                                                                                                                                                                                                                          \\ \hline
15                                                           & SPARQL returns 1 irrelevant result (disambiguation error of the mention {\em \lq{}Betty White'}).                                                                                                                                                                                                                                                                                            \\ \hline
16                                                           & SPARQL returns 2 irrelevant results (disambigation errors of the mention {\em \lq{}butter, chicken'}~  linked to the famous Indian dish Butter Chicken). HistDiv returns 3 relevant results which though are not returned by SPARQL. The results contain the Indian meat dish 'Tandoori Murgh' which is not linked by the entity linking system.       \\ \hline
19                                                           & SPARQL returns 1 irrelevant resutls (disambiguation error of the mention {\em \lq{}Gregory Johnson'}).   \\ \hline
20                                                           & Google returns 1 relevant result which is not returned by SPARQL. The result mentions {\em \lq{}earthquake'} which though is not linked by the entity linking system.                                                                                                                                                                          \\ \hline
\end{tabular}
\end{table*}

\subsubsection*{Efficiency of Query Answering}
The execution time of a SPARQL query over a semantic layer mainly depends on
the following factors:
\begin{itemize}
\item
The efficiency of the triplestore hosting the semantic layer
   (e.g., in-memory triplestores are more efficient).

\item
The efficiency of the server hosting the triplestore (available main memory, etc.).

\item
The query itself since some SPARQL operators are costly (like the operators {\tt FILTER}
and {\tt OPTIONAL}).
Moreover, if the query contains
one or more {\tt SERVICE} operators (like the queries of
Listings \ref{fig:queryExampleIntegrate}-\ref{fig:queryExampleInfer}),
then its execution time is also affected by the efficiency of
the remote endpoints at the time of the request.
\end{itemize}

Table \ref{tbl:exec_time} shows the execution times
of the 20 queries used in our evaluation.
The average execution time was about 400 ms,
with minimum 56 ms for query \#16 and maximum 2.4 sec for query \#15
(we run each query 10 times within 3 days).
All these queries use the {\tt SERVICE} operator for
querying DBpedia's SPARQL endpoint but not any {\tt FILTER} or {\tt OPTIONAL} operator,
while the semantic layer was hosted in a Virtuoso server installed
in a modest personal computer (MacBook Pro, Intel Core i5, 8GB main memory)
and we run the queries in Java 1.8 using Apache Jena 3.1.

\begin{table*}[]
\centering
\caption{Execution times of SPARQL queries.}
\vspace{-1mm}
\setlength{\tabcolsep}{3pt}
\label{tbl:exec_time}
\begin{tabular}{|c|c|c|c|c|c|c|c|c|c|c|c|}
\hline
\rowcolor[HTML]{EFEFEF}
{\color[HTML]{000000} {\bf Query}} & {\color[HTML]{000000} R1 (ms)} & {\color[HTML]{000000} R2 (ms)} & {\color[HTML]{000000} R3 (ms)} & R4 (ms) & R5 (ms) & R6 (ms) & R7(ms) & R8 (ms) & R9 (ms) & R10 (ms) & {\bf Average (ms)} \\ \hline
1 & 40 & 80 & 112 & 78 & 97 & 75 & 74 & 63 & 64 & 63 & 74.6 \\ \hline
2 & 256 & 324 & 406 & 434 & 391 & 408 & 274 & 248 & 251 & 515 & 350.7 \\ \hline
3 & 96 & 585 & 1540 & 532 & 907 & 123 & 120 & 91 & 89 & 97 & 418 \\ \hline
4 & 156 & 169 & 295 & 233 & 176 & 216 & 135 & 137 & 130 & 184 & 183.1 \\ \hline
5 & 55 & 59 & 104 & 98 & 86 & 66 & 168 & 49 & 62 & 71 & 81.8 \\ \hline
6 & 53 & 59 & 67 & 78 & 72 & 67 & 51 & 53 & 189 & 72 & 76.1 \\ \hline
7 & 182 & 199 & 455 & 380 & 497 & 223 & 209 & 182 & 260 & 270 & 285.7 \\ \hline
8 & 58 & 46 & 61 & 103 & 200 & 63 & 43 & 48 & 45 & 110 & 77.7 \\ \hline
9 & 75 & 110 & 181 & 122 & 199 & 126 & 82 & 71 & 67 & 103 & 113.6 \\ \hline
10 & 1,809 & 1,887 & 1,991 & 2,936 & 2,900 & 2,858 & 1,822 & 1,711 & 1,743 & 3,816 & 2,347.3 \\ \hline
11 & 65 & 59 & 60 & 88 & 172 & 81 & 58 & 54 & 62 & 156 & 85.5 \\ \hline
12 & 428 & 431 & 462 & 725 & 883 & 793 & 500 & 399 & 420 & 693 & 573.4 \\ \hline
13 & 42 & 77 & 62 & 96 & 54 & 193 & 40 & 32 & 41 & 50 & 68.7 \\ \hline
14 & 79 & 95 & 92 & 115 & 107 & 407 & 69 & 70 & 62 & 98 & 119.4 \\ \hline
15 & 1,772 & 1,958 & 2,132 & 2,975 & 3,611 & 3,080 & 1,962 & 1,768 & 1,739 & 3,291 & 2,428.8 \\ \hline
16 & 42 & 70 & 74 & 119 & 74 & 45 & 31 & 30 & 33 & 43 & 56.1 \\ \hline
17 & 89 & 95 & 87 & 136 & 117 & 100 & 93 & 79 & 83 & 128 & 100.7 \\ \hline
18 & 182 & 181 & 195 & 258 & 581 & 253 & 235 & 153 & 162 & 229 & 242.9 \\ \hline
19 & 65 & 98 & 71 & 100 & 712 & 81 & 89 & 55 & 51 & 85 & 140.7 \\ \hline
20 & 82 & 81 & 67 & 108 & 104 & 72 & 56 & 57 & 61 & 88 & 77.6 \\ \hline
\multicolumn{11}{|r|}{\cellcolor[HTML]{EFEFEF}{\bf Average (ms):}} & {\bf 395.12} \\ \hline
\end{tabular}
\end{table*}

\section{Conclusion}
\label{sec:concl}

We have introduced a model and a framework
for describing and publishing metadata and
semantic information about web archives.
The constructed {\em semantic layers} allow:
i) exploring web archives in a more advanced way
based on entities, events and concepts extracted
from the archived documents and linked to web resources;
ii) integrating information (even at query-execution time)
coming from multiple knowledge bases and semantic layers;
iii) inferring new knowledge that is very laborious to derive otherwise;
iv) coping with common problems when exploring web archives like
temporal reference variants and multilinguality; and
v) making the contents of web archives machine understandable,
thereby enabling their direct exploitation by other systems and tools.
The results of a comparative evaluation showed that semantic layers
can answer complex information needs that keyword-based search systems
fail to sufficiently satisfy.
The evaluation also enabled us to identify problems that can
affect the effectiveness of query answering.

We believe that constructing semantic layers is
the first step towards more advanced and meaningful
exploration of web archives \cite{holzmannaccessing2017}.
Our vision is to enrich the LOD cloud\footnote{\url{http://lod-cloud.net/}}
with semantic layers, i.e., with knowledge bases describing metadata
and semantic information about archived collections.

Regarding future work and research,
user-friendly interfaces should be
developed on top of semantic layers for allowing
end-users to easily and efficiently explore web archives.
Another interesting direction is to study approaches
for ranking the results returned by SPARQL queries \cite{fafalios2017jcdlPoster}.

\begin{acknowledgements}
The work was partially funded by the
European Commission for the ERC Advanced Grant ALEXANDRIA (No. 339233).
\end{acknowledgements}

\bibliographystyle{spbasic}
\bibliography{SemLayer_IJDL_FINAL}

\end{document}